\documentclass[12pt]{article}
\usepackage{here,amsfonts,pstricks,pst-node,
pst-plot,pst-coil,epsfig,psfig}
\newbox\sample
\newif\ifproofmode
\newif\ifsymindex
\global\symindexfalse
\newwrite\inx
\def\indsyma#1#2{\ifproofmode\marginpar{$\scriptstyle#1$}\fi%
\ifx#2\empty\write\inx{$\noexpand#1$,\space\thepage}%
\write\inx{\string\newline}\else%
\write\inx{$\noexpand#1$,\space#2,\space\thepage}%
\write\inx{\string\newline}\fi\ignorespaces}%
\def\indsym#1#2{\ifsymindex%
\ifproofmode\marginpar{$\scriptstyle#1$}\fi%
\ifx#2\empty\write\inx{\string\item \space$\noexpand#1$,\space\thepage}%
\else%
\write\inx{\string\item \space$\noexpand#1$,\space#2,\space\thepage}%
\fi\ignorespaces\fi}%
\newskip\dangerskipb
\newskip\dangerskip
\def\hang{\hangindent\dangerskip}

\def\s#1{{\cal #1}}

\def\remarks{\bigskip\noindent{\bf Remarks:}\enspace}

\font\manual=manfnt at 12pt
\def\danbend{{\manual\char127}}
\def\ddatanger{\medbreak\begingroup\clubpenalty=10000
 \def\par{\endgraf\endgroup\medbreak} \noindent\hang\hangafter=-2
 \hbox to0pt{\hskip-3.5pc\danbend\kern1pt%
\danbend\hfill}}

\def\dobdownarrow{\mathop{\vbox{\kern2pt \hbox{$\Big\downarrow$}\kern-16.5pt
                          \nointerlineskip\hbox{$\Big\downarrow$}}}}

\def\lrightarrow{\hbox to 25pt{\rightarrowfill}}

\def\supexp{exp(m,n,p)=m^{m^{m^{\cdot^{\cdot^{\cdot^{m^{p}}}}}}}
\vbox{\hbox{$\Big\}\scriptstyle n$}\kern0pt}}

\def\supexpo#1#2#3{#1^{#1^{\cdot^{\cdot^{\cdot^{#1^{#2}}}}}}
\vbox{\hbox{$\Big\}\scriptstyle #3$}\kern0pt}}

\def\sqr#1#2{{\vcenter{\hrule height .#2pt
         \hbox{\vrule width.#2pt height#1pt \kern#1pt
             \vrule width.#2pt}
         \hrule height.#2pt}}}

\def\co{\colon}

\newskip\bogcentering \bogcentering= 0pt plus 1000pt minus 1000pt 

\def\matth{\mathsurround=0pt}
\def\fakrightarrowfill{$\matth \mathord- \mkern-6mu
  \cleaders\hbox{$\mkern-2mu \mathord- \mkern-2mu$}\hfill
 \mkern-6mu \mathord\rightarrow$}

\def\fakoverrightarrow#1{\vbox{\ialign{##\crcr
  \fakrightarrowfill\crcr\noalign{\kern-1pt\nointerlineskip}
 $\hfil\displaystyle{#1}\hfil$\crcr}}}

\newif\ifdtatp

\def\displaty{%
\global \dtatptrue \openup \jot \matth \everycr{\noalign{\ifdtatp \global 
\dtatpfalse \vskip -\lineskiplimit \vskip \normallineskiplimit \else 
\penalty \interdisplaylinepenalty \fi }}}

\def\displaylignes#1{\displaty
   \halign{\hbox to\displaywidth{$\displaystyle##$}\crcr
   #1\crcr}}
\def\eqaligneno#1{\displaty \tabskip=\bogcentering
 \halign to\displaywidth{\hfil$\displaystyle{##}$\tabskip=0pt
 &$\displaystyle{{}##}$\hfil\tabskip=\bogcentering
 &\llap{$##$}\tabskip=0pt\crcr
 #1\crcr}}
\def\leqaligneno#1{\displaty \tabskip=\bogcentering
 \halign to\displaywidth{\hfil$\displaystyle{##}$\tabskip=0pt
 &$\displaystyle{{}##}$\hfil\tabskip=\bogcentering
 &\kern-\displaywidth\rlap{$##$}\tabskip=\displaywidthpt\crcr
 #1\crcr}}
\def\ligne{\hbox to\hsize}
\newdimen\nouvpagewidth
\newdimen\offwidth
\newdimen\lawidthoui
\nouvpagewidth=\lawidthoui
\def\kboxit#1{\vbox{\hrule\hbox{\vrule\kern3pt
              \vbox{\kern3pt#1\kern3pt}\kern3pt\vrule}\hrule}}

\def\kboxitb#1{\vbox{\hrule\hbox{\vrule\kern3pt
              \vbox{\kern3pt#1\kern3pt}\kern3pt\vrule}\hrule}}

\def\laboxaround#1{
\aboxaround{\hbox to\hsize{\hfill\box2\hfill}}{#1}
}

\def\boxar#1#2{
\aboxaround{\hbox to\hsize{\hfill#1\hfill}}{#2}
}

\def\aboxaround#1#2{
\setbox4=\vbox{\hsize #2\noindent\strut#1\strut}
\kboxitb{\box4}}

\def\kframeit#1{\vbox{\hrule\hbox{\vrule\kern5pt
              \vbox{\kern5pt#1\kern5pt}\kern5pt\vrule}\hrule}}
\newskip\savnormalbaselineskip
\newskip\savnormallineskip
\newdimen\savnormallineskiplimit

\def\matrice#1{
\savnormalbaselineskip=\normalbaselineskip
\savnormallineskip=\normallineskip
\savnormallineskiplimit=\normallineskiplimit
\normalbaselineskip=16pt
\normallineskip=2pt \normallineskiplimit2pt
\matrix{#1}
\normalbaselineskip=\savnormalbaselineskip
\normallineskip=\savnormallineskip
\normallineskiplimit=\savnormallineskiplimit}

\newtheorem{thm}{Theorem}[section]
\newtheorem{lemma}[thm]{Lemma}

\def\ringpoly#1#2{#1[#2]}
\def\mdeg{m}
\def\ndeg{m}

\def\reals{\mathbb{R}}

\def\mapdef#1#2#3{#1\co #2\rightarrow #3}

\def\affs{\s{E}}
\def\card{\mathrm{card}}
\def\natnums{\mathbb{N}}
\def\batop#1#2{{\displaystyle #1\atop \displaystyle #2}}

\begin{document}
\title{On the Efficiency of Strategies for  Subdividing
Polynomial Triangular Surface Patches}
\author{Jean Gallier\\
%\author{Elie Cartan, Guido Castelnuovo,\\
%Eduardo Clog, and Kurt W.A.J.H.Y. Reillag\\
%Leonardo M.A.D. Ed Olrac, and Kurt W.A.J.H.Y. Reillag\\
 \\
Department of Computer and Information Science\\
University of Pennsylvania\\
Philadelphia, PA 19104, USA\\ \\
%{\tt cartan@sorbonne.vin.paris.edu}\\
%{\tt castelnuovo@vesuvio.parmegiano.roma.edu}\\
{\tt jean@saul.cis.upenn.edu}\\
}
\maketitle
\vspace{0.3cm}
\noindent
{\bf Abstract.}
In this paper, we investigate the efficiency of various 
strategies for subdividing polynomial triangular surface patches. 
We give a simple algorithm performing a regular subdivision
in four calls to the standard
de Casteljau algorithm (in its subdivision version).
A naive version uses twelve calls.
We also show that any method for obtaining a regular subdivision
using the standard de Casteljau algorithm requires at least $4$ calls.
Thus, our method is optimal. 
We give another  subdivision algorithm using only three calls 
to the de Casteljau algorithm. Instead of being regular,
the subdivision pattern is  diamond-like. Finally, we present a
``spider-like'' subdivision scheme producing six subtriangles in
four calls to the de Casteljau algorithm.

%\vfill\eject
%\tableofcontents
\vfill\eject
\section{Introduction}
\label{sec1}
In this paper, we investigate the efficiency of various 
strategies for subdividing polynomial triangular surface patches. 
Subdivision methods based on a version of the de Casteljau algorithm
splitting a control net into control subnets (see Farin \cite{Farin86})
were investigated by Goldman \cite{Goldman83}, Boehm and Farin
\cite{Bofa83},  B\"ohm \cite{Boehm83b}, 
and Seidel \cite{Seidel89a}
(see also Boehm, Farin, and Kahman \cite{Boehm84}, and Filip \cite{Filip86}).
However, except for B\"ohm \cite{Boehm83b}, 
these papers are not particularly concerned with
minimizing the number of calls to the standard de Casteljau algorithm.
Furthermore, some of these papers (notably Goldman \cite{Goldman83})
use a version of the  de Casteljau algorithm
computing a $5$-dimensional simplex of polar values, which
is more expensive than the standard $3$-dimensional version.
In this paper, 
we give a simple algorithm performing a regular subdivision
in four calls to the standard de Casteljau algorithm (in its subdivision version).
A naive version uses twelve calls.
We also show that any method for obtaining a regular subdivision
using the standard de Casteljau algorithm requires at least $4$ calls.
Thus, our method is optimal. 
We give another  subdivision algorithm using only three calls 
to the de Casteljau algorithm. Instead of being regular,
the subdivision pattern is  diamond-like. Finally, we present a
``spider-like'' subdivision scheme producing six subtriangles in
four calls to the de Casteljau algorithm.
Some familiarity with affine spaces and affine maps is assumed.
Details can be found in Farin \cite{Farin93}, Berger \cite{Berger90}, or
Gallier \cite{Gallbook}.
\section{The Polar Form Approach to Polynomial Triangular Surface Patches}
\label{sec2}
The deep reason why polynomial triangular surface patches can be
effectively handled in terms of control points is that
multivariate polynomials arise from multiaffine symmetric maps
(see Ramshaw \cite{Ramshaw87}, Farin \cite{Farin93,Farin95}, 
Hoschek and Lasser \cite{Hoschek}, 
or Gallier \cite{Gallbook}).
Denoting the affine plane $\reals^2$ as $\s{P}$, 
traditionally, a  {\it polynomial surface\/} in $\reals^n$ is a function
$\mapdef{F}{\s{P}}{\reals^n}$, defined such that
$$\eqaligneno{
x_1 &= F_1(u,v),\cr
\ldots &= \ldots\cr
x_n &= F_n(u,v),\cr
}$$
for all $u, v\in\reals$,
where $F_1(U,V),\ldots, F_{\ndeg}(U, V)$ are polynomials 
in $\ringpoly{\reals}{U, V}$. 
Given a natural number $\mdeg$,
if each polynomial $F_i(U, V)$ has total degree $\leq \mdeg$,
we say that $F$ is a {\it polynomial surface of total degree $\mdeg$\/}.
The {\it trace of the surface $F$\/} is the set $F(\s{P})$.

\medskip
Now, given  a polynomial surface $F$ of total degree $\mdeg$
in some affine space $\affs$ (typically $\reals^3$),
there is unique symmetric and multiaffine map
$\mapdef{f}{\s{P}^{\mdeg}}{\affs}$
such that
$$F(u,v) = f(\underbrace{(u, v),\ldots,(u, v)}_{\mdeg}),$$
for all $(u, v)\in\s{P}$. The symmetric and multiaffine map $f$ associated
with $F$ is called the {\it polar form of $F$\/}.

\medskip
The above result is not hard to prove. Using linearity, it is
enough to deal with a single monomial.
Given a monomial $U^hV^k$, 
with $h + k = d\leq \mdeg$, it is easily shown that
the symmetric multiaffine $f$ form corresponding to $U^hV^k$ is given
$$f((u_1, v_1), \ldots, (u_\mdeg, v_\mdeg)) =
\frac{h!k!(m - (h + k))!}{m!}\>
\sum_{\scriptstyle I\cup J \subseteq \{1,\ldots,\mdeg\} \atop
{\scriptstyle I\cap J = \emptyset \atop
\scriptstyle \card{(I)} = h,\, \card{(J)} = k}}
\biggl(\prod_{i\in I} u_i\biggr) \biggl(\prod_{j\in J} v_j\biggr).$$

\medskip
Recall that a map $\mapdef{f}{\reals^d}{\reals^n}$ is {\it affine\/} if
$$f((1 - \lambda) a + \lambda b) = (1 - \lambda) f(a) + \lambda f(b),$$
for all $a, b\in \reals^d$, and all $\lambda\in\reals$.
A map $\mapdef{f}{\underbrace{\reals^d\times \cdots\times \reals^d}_{m}}{\reals^n}$ 
is {\it multiaffine\/} if it is affine in each of its arguments.
A map  $\mapdef{f}{\underbrace{\reals^d\times \cdots\times \reals^d}_{m}}{\reals^n}$ 
is {\it symmetric\/} if it does
not depend on the order of its arguments, i.e.,
$f(a_{\pi(1)},\ldots, a_{\pi(m)}) = f(a_1,\ldots,a_m)$,
for all $a_1, \ldots, a_m$, and all permutations $\pi$.

\medskip
As an example, consider the following surface known as
Enneper's surface:
$$\eqaligneno{
F_1(U, V) &= U - \frac{U^3}{3} + UV^2\cr
F_2(U, V) &= V - \frac{V^3}{3} + U^2V\cr
F_3(U, V) &= U^2 - V^2.\cr
}$$

\medskip
We get the polar forms
$$\eqaligneno{
f_1((U_1, V_1), (U_2, V_2), (U_3, V_3)) 
&= \frac{U_1 + U_2 + U_3}{3}  - \frac{U_1U_2U_3}{3}\cr
&\qquad
+ \frac{U_1V_2V_3 + U_2V_1V_3 + U_3V_1V_2}{3}\cr
f_2((U_1, V_1), (U_2, V_2), (U_3, V_3)) 
&= \frac{V_1 + V_2 + V_3}{3} - \frac{V_1V_2V_3}{3}\cr
&\qquad
+ \frac{U_1U_2V_3 + U_1U_3V_2 + U_2U_3V_1}{3}\cr
f_3((U_1, V_1), (U_2, V_2), (U_3, V_3)) 
&= \frac{U_1U_2 + U_1U_3 + U_2U_3}{3} - \frac{V_1V_2 + V_1V_3 + V_2V_3}{3}.\cr
}$$

\medskip
Furthermore, it turns out that any symmetric multiaffine map
$\mapdef{f}{\s{P}^{\mdeg}}{\affs}$ is uniquely determined
by a family of $\frac{(\mdeg + 1)(\mdeg + 2)}{2}$ points
(where $\affs$ is any affine space, say $\reals^n$).
Let
$$\Delta_{\mdeg} = \{(i,\, j,\, k)\in\natnums^3
\ |\ i + j + k = \mdeg\}.$$
The following lemma is easily shown (see Ramshaw \cite{Ramshaw87} or
Gallier \cite{Gallbook}).

\begin{lemma}
\label{topolslem1}
Given an affine frame  $\Delta rst$ in the plane
$\s{P}$, given a family
$(b_{i,\,j,\,k})_{(i,j,k)\in\Delta_{\mdeg}}$ of 
$\frac{(\mdeg + 1)(\mdeg + 2)}{2}$ points in $\affs$, 
there is a unique surface $\mapdef{F}{\s{P}}{\affs}$
of total degree $\mdeg$,
defined by a symmetric $\mdeg$-affine polar form
$\mapdef{f}{\s{P}^{\mdeg}}{\affs}$, such that
$$f(\underbrace{r,\ldots,r}_{i}, 
\underbrace{s,\ldots,s}_{j},
\underbrace{t,\ldots,t}_{k}) = b_{i,\,j,\,k}$$
for all $(i,j,k)\in\Delta_{\mdeg}$.
Furthermore, $f$ is given by the expression
$$f(a_1,\ldots,a_\mdeg) =
\sum_{\scriptstyle I\cup J\cup K = \{1,\ldots,\mdeg\}  \atop
      \scriptstyle I, J, K\ disjoint
     }
\biggl(\prod_{i\in I} \lambda_i\biggr) \biggl(\prod_{j\in J} \mu_j\biggr) 
\biggl(\prod_{k\in K} \nu_k\biggr) \>
f(\underbrace{r,\ldots,r}_{\card{(I)}}, 
\underbrace{s,\ldots,s}_{\card{(J)}},
\underbrace{t,\ldots,t}_{\card{(K)}}),$$
where  $a_i = \lambda_i r + \mu_i s + \nu_i t$, 
with $\lambda_i + \mu_i + \nu_i = 1$, and $1\leq i\leq \mdeg$.
\end{lemma}

\medskip
For example, with respect to the standard frame
$\Delta rst = ((1, 0, 0), (0, 1, 0), (0, 0, 1))$, we obtain the
following $10$ control  points for the Enneper surface:

\medskip
$$\matrice{
&  &  & \batop{f(r,r,r)}{(\frac{2}{3},0,1)}    &  &  &\cr
 &  &\batop{f(r,r,t)}{(\frac{2}{3},0,\frac{1}{3})} &  &
\batop{f(r,r,s)}{(\frac{2}{3},\frac{2}{3},\frac{1}{3})} &  &\cr
   &\batop{f(r,t,t)}{(\frac{1}{3},0,0)} &  
&\batop{f(r,s,t)}{(\frac{1}{3},\frac{1}{3},0)} & 
& \batop{f(r,s,s)}{(\frac{2}{3},\frac{2}{3},-\frac{1}{3})} &\cr
\batop{f(t,t,t)}{(0,0,0)} &    &  \batop{f(s,t,t)}{(0,\frac{1}{3},0)} &     & 
\batop{f(s,s,t)}{(0,\frac{2}{3},-\frac{1}{3})} &       &
\batop{f(s,s,s)}{(0,\frac{2}{3},-1)} \cr
}$$

\medskip
A family  $\s{N}= (b_{i,\,j,\,k})_{(i,j,k)\in\Delta_{\mdeg}}$ of 
$\frac{(\mdeg + 1)(\mdeg + 2)}{2}$ points in $\affs$ is called
a {\it (triangular) control net, or B\'ezier net\/}.
Note that the points in
$$\Delta_{\mdeg} = \{(i,\, j,\, k)\in\natnums^3
\ |\ i + j + k = \mdeg\},$$
can be thought of as a triangular grid of points in $\s{P}$.
For example, when $\mdeg = 5$, we have the following grid of
$21$ points:

\medskip
$$\matrice{
    &     &     &     &     & 500 &     &     &     &     &     \cr
    &     &     &     & 401 &     & 410 &     &     &     &     \cr
    &     &     & 302 &     & 311 &     & 320 &     &     &     \cr
    &     & 203 &     & 212 &     & 221 &     & 230 &     &     \cr
    & 104 &     & 113 &     & 122 &     & 131 &     & 140 &     \cr
005 &     & 014 &     & 023 &     & 032 &     & 041 &     & 050 \cr
}$$

\bigskip
We intentionally let $i$ be the row index, starting from the 
left lower corner, and $j$ be the column index, 
also starting from the left lower corner.
The control net  $\s{N}= (b_{i,\,j,\,k})_{(i,j,k)\in\Delta_{\mdeg}}$ 
can be viewed as an image of the triangular grid $\Delta_{\mdeg}$ in
the affine space $\affs$.
It follows from Lemma \ref{topolslem1} that there is a bijection between
polynomial surfaces of degree $\mdeg$ and control nets 
$\s{N}= (b_{i,\,j,\,k})_{(i,j,k)\in\Delta_{\mdeg}}$.
It should also be noted that there are efficient methods for
computing control nets from parametric definitions,
but this will be published elsewhere.

\medskip
In the next section, we review a beautiful algorithm
to compute a point $F(a)$ on a surface patch using
affine interpolation steps, the de Casteljau algorithm.

\section{The de Casteljau Algorithm for Triangular Patches}
\label{sec3}
In this section, we explain in detail how the de Casteljau
algorithm can be used to subdivide a triangular patch
into three subpatches. For more details, 
see Farin \cite{Farin93,Farin95},
Hoschek and Lasser \cite{Hoschek}, Risler \cite{Risler92},
or Gallier \cite{Gallbook}.
In the next section, we will use versions of this algorithm
to obtain a triangulation of a surface
patch using recursive subdivision.

\medskip
Given an affine frame $\Delta rst$, 
given a triangular control net  
$\s{N}= (b_{i,\,j,\,k})_{(i,j,k)\in\Delta_{\mdeg}}$,
recall that in terms of the polar form $\mapdef{f}{\s{P}^{\mdeg}}{\affs}$
of the polynomial surface $\mapdef{F}{\s{P}}{\affs}$
defined by $\s{N}$, for every $(i, j, k)\in \Delta_{\mdeg}$,
we have
$$b_{i,\, j,\, k} = 
f(\underbrace{r,\ldots,r}_{i}, \underbrace{s,\ldots,s}_{j},
\underbrace{t,\ldots,t}_{k}).$$
Given $a = \lambda r + \mu s + \nu t$ in $\s{P}$, 
where $\lambda + \mu + \nu = 1$,
in order to compute $F(a) = f(a, \ldots, a)$,
the computation builds a sort of tetrahedron consisting
of $\mdeg + 1$ layers. The base layer consists of
the original control points in $\s{N}$, which are also denoted as
$(b^{0}_{i,\,j,\,k})_{(i,j,k)\in\Delta_{\mdeg}}$.
The other layers are computed in
$\mdeg$ stages, where at stage $l$, $1\leq l \leq \mdeg$, 
the points $(b^{l}_{i,\,j,\,k})_{(i,j,k)\in\Delta_{\mdeg - l}}$
are computed such that
$$b^{l}_{i,\, j,\, k} = \lambda b^{l-1}_{i+1,\, j,\, k} 
+ \mu b^{l-1}_{i,\, j + 1,\, k} + \nu b^{l-1}_{i,\, j,\, k+1}.$$ 
During the last stage, the single
point $b^{\mdeg}_{0,\, 0,\, 0}$ is computed.
An easy induction shows that
$$b^{l}_{i,\, j,\, k} = 
f(\underbrace{a,\ldots,a}_{l}, \underbrace{r,\ldots,r}_{i}, 
\underbrace{s,\ldots,s}_{j},
\underbrace{t,\ldots,t}_{k}),$$
where $(i,j,k)\in\Delta_{\mdeg - l}$,
and thus, $F(a) = b^{\mdeg}_{0,\, 0,\, 0}$.

\medskip
Assuming that $a$ is not on one of the edges of
$\Delta rst$,
the crux of the subdivision method is that
the three other faces of the tetrahedron of polar values
$b^{l}_{i,\, j,\, k}$ besides the face corresponding to the original
control net, yield three control nets
$$\s{N}{ast} = (b^{l}_{0,\, j,\, k})_{(l, j, k)\in \in\Delta_{\mdeg}},$$
corresponding to the base triangle $\Delta ast$,
$$\s{N}{rat} = (b^{l}_{i,\, 0,\, k})_{(i, l, k)\in \in\Delta_{\mdeg}},$$
corresponding to the base triangle $\Delta rat$,
and
$$\s{N}{rsa} = (b^{l}_{i,\, j,\, 0})_{(i, j, l)\in \in\Delta_{\mdeg}},$$
corresponding to the base triangle $\Delta rsa$.
If $a$ belongs to one of the edges, say $rs$, then the triangle
$\Delta rsa$ is flat, i.e. $\Delta rsa$ is not an afine frame,
and the net $\s{N}{rsa}$ does not define the surface, but instead a curve.
However, in such cases, the degenerate net $\s{N}{rsa}$ is not needed anyway.

\medskip
From an implementation point of view, we found it convenient
to assume that a triangular net
$\s{N}= (b_{i,\,j,\,k})_{(i,j,k)\in\Delta_{\mdeg}}$
is represented as the list consisting of the concatenation
of the $\mdeg + 1$ rows 
$$b_{i,\,0,\,\mdeg - i},\> b_{i,\,1,\,\mdeg - i - 1},\> \ldots,\>
b_{i,\,\mdeg - i,\,0},$$ 
i.e.,
$$f(\underbrace{r,\ldots, r}_{i},\>
\underbrace{t,\ldots, t}_{\mdeg - i}),\>
f(\underbrace{r,\ldots, r}_{i},\> s,\>
\underbrace{t,\ldots, t}_{\mdeg - i - 1}),
\> \ldots, \>    
f(\underbrace{r,\ldots, r}_{i},\>
\underbrace{s,\ldots, s}_{\mdeg - i - 1},\> t), \>
f(\underbrace{r,\ldots, r}_{i},\>
\underbrace{s,\ldots, s}_{\mdeg - i}),$$    
where $0 \leq i \leq \mdeg$.
As a triangle, the net $\s{N}$ is listed (from top-down) as

$$\displaylignes{
\hfill f(\underbrace{t,\ldots, t}_{\mdeg})\quad
f(\underbrace{t,\ldots, t}_{\mdeg - 1},\, s)\quad \ldots \quad    
f(t, \>\underbrace{s,\ldots, s}_{\mdeg - 1}) \quad
f(\underbrace{s,\ldots, s}_{\mdeg})    
\hfill\cr
\hfill \ldots\qquad\qquad\qquad\qquad \ldots \hfill\cr
\hfill \ldots \hfill\cr
\hfill f(\underbrace{r,\ldots, r}_{\mdeg - 1},\, t)\quad    
f(\underbrace{r,\ldots, r}_{\mdeg - 1},\> s)  \hfill\cr  
\hfill f(\underbrace{r,\ldots, r}_{\mdeg})    
\hfill\cr
}$$

\bigskip
The main advantage of this representation is that
we can view the net $\s{N}$ as a two-dimensional array {\it net\/},
such that $\hbox{\it net\/}[i, j] = b_{i,\, j,\, k}$
(with $i + j + k = \mdeg$). In fact, only a triangular
portion of this array is filled.
This way of representing control nets fits well with
the convention that the affine frame $\Delta rst$
is represented as follows:

\medskip

\begin{figure}[H]
%\begin{example}
  \begin{center}
    \begin{pspicture}(0,-0.5)(6,4)
    \psline[linewidth=1.5pt](0,0)(3,4)
    \psline[linewidth=1.5pt](0,0)(6,0)
    \psline[linewidth=1.5pt](6,0)(3,4)
    \psline[linewidth=1.5pt](3,1.33)(0,0)
    \psline[linewidth=1.5pt](3,1.33)(6,0)
    \psline[linewidth=1.5pt](3,1.33)(3,4)
    \uput[-135](0,0){$t$}
    \uput[-45](6,0){$s$}
    \uput[90](3,4){$r$}
    \uput[45](3,1.33){$a$}
    \end{pspicture}
  \end{center}
  \caption{An affine frame}
%\end{example}
\end{figure}
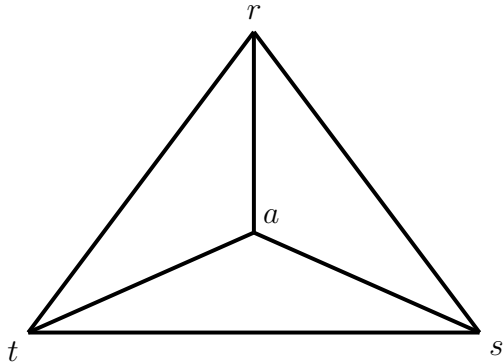

\medskip
Instead of simply computing $F(a) = b^{\mdeg}_{0,\, 0,\, 0}$, the
de Casteljau algorithm can be easily adapted to
output the three nets $\s{N}{ast}$, $\s{N}{rat}$, and
$\s{N}{rsa}$. The function  {\tt sdecas3\/} does that.
We also found it convenient to write three distinct functions
{\tt subdecas3ra\/}, {\tt subdecas3sa\/}, and
{\tt subdecas3ta\/}, computing the control nets with respect to the
affine frames $\Delta ast$, $\Delta art$, and $\Delta ars$.
An implementation in {\it Mathematica\/} can be found in
Gallier \cite{Gallbook}.

\section{Regular Subdivision Of Triangular Patches}
\label{sec4}
If we want to render a triangular surface patch $F$ defined
over the affine frame $\Delta rst$, it seems natural
to subdivide  $\Delta rst$ into the three subtriangles
$\Delta ars$, $\Delta ast$, and $\Delta art$, where $a = (1/3, 1/3, 1/3)$ 
is the center of gravity of the triangle $\Delta rst$, getting
new control nets $\s{N}ars$, $\s{N}ast$ and $\s{N}art$
using the functions described earlier, and repeat this process
recursively. However, this process does not yield a good 
triangulation of the surface patch, because no progress is made
on the edges $rs$, $st$, and $tr$, and thus,
such a triangulation does not converge to the surface patch.
Thus, in order to compute triangulations that converge to the
surface patch, we need to subdivide the triangle
$\Delta rst$ in such a way that the edges of the affine
frame are subdivided. There are many ways of performing
such subdivisions, and we will propose a method which has
the advantage of yielding a very regular triangulation, and
of being very efficient. In fact, we give an optimal  method
for subdividing an affine frame using four
calls to the standard de Casteljau algorithm in its subdivision version.
A naive method would require twelve calls.

\medskip
Goldman \cite{Goldman83} proposed several subdivision
algorithms, including one for splitting a triangular patch
into four triangular subpatches, but his methods use a generalized
version of the de Casteljau algorithm computing a $5$-simplex 
of polar values. These methods are illustrated graphically 
in Boehm and Farin \cite{Bofa83}.
It should be noted that Boehm and Farin do mention that
it is possible to compute the control net w.r.t. a new affine
frame from the control  net w.r.t. an original affine frame in three
calls to the standard de Casteljau algorithm. However, they do not
explain how to split a triangular patch into four subpatches
using four calls to the standard de Casteljau algorithm.

\medskip
Goldman's subdivision methods can be justified in a very simple way
as shown by Seidel \cite{Seidel89a}.
Given a surface $F$ of total degree $\mdeg$ defined by a
triangular control net
$\s{N}= (b_{i,\,j,\,k})_{(i,j,k)\in\Delta_{\mdeg}}$,
w.r.t. the affine frame $\Delta rst$,
for any $n$ points $p_i = u_i r + v_i s + w_i t$
(where $u_i + v_i + w_i = 1$), the following
$(n + 2)$-simplex of points $b_{i, j, k}^{l_1,\ldots,l_n}$ 
where $i + j + k + l_1 + \ldots + l_n = \mdeg$ is defined inductively
as follows:
$$\eqaligneno{
b_{i, j, k}^{0,\ldots,0} &= b_{i, j, k},\cr
b_{i, j, k}^{l_1,\ldots,l_{h+1},\ldots,l_n} &= 
u_h\,b_{i+1, j, k}^{l_1,\ldots,l_h,\ldots,l_n} 
+ v_h\,b_{i, j+1, k}^{l_1,\ldots,l_h,\ldots,l_n} 
+ w_h\,b_{i, j, k+1}^{l_1,\ldots,l_h,\ldots,l_n},\cr
}$$
where $1\leq h \leq n$.

\medskip
If $f$ is the polar form of $F$, it is easily shown that
$$b_{i, j, k}^{l_1,\ldots,l_n} = 
f(\underbrace{r,\ldots,r}_i, \underbrace{s,\ldots,s}_j,
\underbrace{t,\ldots,t}_k, \underbrace{p_1,\ldots,p_1}_{l_{1}}, \ldots,
\underbrace{p_n,\ldots,p_n}_{l_{n}}).$$

\medskip
For $n = 0$, $F(p) = b^n_{0,0,0}$,
as in the standard de Casteljau algorithm.
For $n=3$,
$(b^{l_{1},l_{2},l_{3}}_{0,\,0,\,0})_{(l_1,l_2,l_2)\in\Delta_{\mdeg}}$
is a control net of $F$ w.r.t. $\Delta p_1p_2p_3$.

\medskip
In particular, if $p_1, p_2, p_3$ are chosen on the edges of $\Delta rst$,
the subnets for the four subpatches  are obtained.
Goldman observes that some of the nets involved in the computation are
trivial, but still, a $5$-simplex of polar values is computed.

\medskip
It was brought to our attention by Gerald Farin (and it is  mentioned
in Remark 2 of Seidel's paper  \cite{Seidel89a}, page 580) that
Helmut Prautzsch showed in his dissertation (in German) \cite{Prautzsch84} 
that regular subdivision into four subtriangles can be achieved
in four calls to the standard de Casteljau algorithm.
Prautzsch's method is briefly described in
B\"ohm \cite{Boehm83b}, page 348 (figure) and page 349 
(in fact, with a typo, one of the
barycentric coordinates listed as
$\left(\frac{1}{2},\frac{1}{2}, 1\right)$ should be
$\left(\frac{1}{2},\frac{1}{2}, 0\right)$).
The order in which the four patches are obtained is slightly
different from ours.
Since Prautzsch's algorithm has not
been discussed more extensively in the literature,
we feel justified in presenting our  method.

\medskip
The subdivision strategy that we will follow is to divide
the affine frame $\Delta rst$ into four subtriangles
$\Delta abt$, $\Delta bac$, $\Delta crb$, and $\Delta sca$,
where $a = (0, 1/2, 1/2)$, 
$b = (1/2, 0, 1/2)$, and $c = (1/2, 1/2, 0),$ 
are the middle points of the sides $st$,
$rt$ and $rs$ respectively, as shown in the diagram below:

\medskip

\begin{figure}[H]
%\begin{example}
  \begin{center}
    \begin{pspicture}(0,-0.5)(8,6)
    \psline[linewidth=1.5pt](0,0)(4,6)
    \psline[linewidth=1.5pt](0,0)(8,0)
    \psline[linewidth=1.5pt](8,0)(4,6)
    \psline[linewidth=1.5pt](2,3)(6,3)
    \psline[linewidth=1.5pt](2,3)(4,0)
    \psline[linewidth=1.5pt](6,3)(4,0)
    \uput[-135](0,0){$t$}
    \uput[-45](8,0){$r$}
    \uput[90](4,6){$s$}
    \uput[135](2,3){$a$}
    \uput[-90](4,0){$b$}
    \uput[45](6,3){$c$}
    \uput[90](2,1){$abt$}
    \uput[90](4,1.5){$bac$}
    \uput[90](6,1){$crb$}
    \uput[90](4,4){$sca$}
    \end{pspicture}
  \end{center}
  \caption{Subdividing an affine frame $\Delta rst$}
%\end{example}
\end{figure}
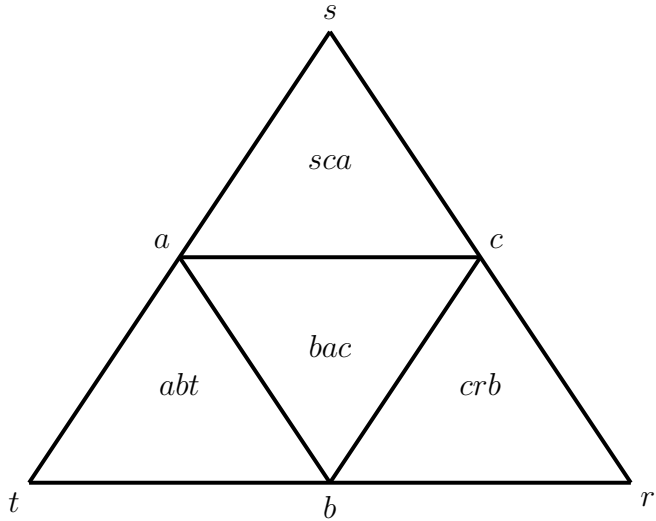

\medskip
First, we show that any method using the standard version of the
de Casteljau algorithm for subdividing a triangular patch
into $4$ subpatches forming a regular pattern as above requires
$4$ calls. The crux of the argument
is that a call to the de Casteljau
algorithm in its subdivision version produces three subpatches
containing only one new corner. We want to produce
the four subpatches $abt$, $crb$, $sca$, and $abc$.
After one subdivision step, we have three
patches each involving exactly one of $a, b, c$. 
After two subdivision steps, we have
six subspatches only two of which involve exactly two of $a, b, c$,
since we can only subdivide a single patch, 
and since this patch only has one of $a, b, c$.
Thus, at least three steps are needed to produce
four subpatches involving at least  two of   $a, b, c$.
If we produced the patch $abc$ during the third subdivision step,
we would have three patches involving exactly two of $a, b, c$, 
but the subdivision step that produced $abc$ also produces
two patches sharing the same vertex from $(r, s, t)$.
However, $abt$, $crb$, and $sca$ do not share a
vertex from $(r, s, t)$. If $abc$ was not produced during the third
step,  at least four steps are needed. Therefore,
in all cases, at least four steps are needed
to produce the required four subpatches.

\medskip
We now present our algorithm.
The first step is to compute the control net for the
affine frame $\Delta bat$. This can be done using two
steps. In the first step, split the triangle $\Delta rst$
into the two triangles $\Delta art$ and $\Delta ars$,
where $a = (0, 1/2, 1/2)$ is the middle of $st$.
Using the function {\tt sdecas3\/} (with $a = (0, 1/2, 1/2)$), the nets
$\s{N}art$, $\s{N}ast$, and $\s{N}ars$ are obtained, and we throw away
$\s{N}ast$ (which is degenerate anyway).
Then, we split $\Delta art$ into
the two triangles $\Delta bat$ and $\Delta bar$. For this, we need
the barycentric coordinates of $b$ with respect to the triangle 
$\Delta art$, which turns out $(0, 1/2, 1/2)$.
Using the function {\tt sdecas3\/}, the nets
$\s{N}bat$, $\s{N}brt$, and $\s{N}bar$ are obtained, and we throw away
$\s{N}brt$.

\medskip

\begin{figure}[H]
%\begin{example}
  \begin{center}
    \begin{pspicture}(0,-0.5)(8,6)
    \psline[linewidth=1pt](0,0)(4,6)
    \psline[linewidth=1pt](0,0)(8,0)
    \psline[linewidth=1pt](8,0)(4,6)
    \psline[linewidth=1.5pt](2,3)(8,0)
    \psline[linewidth=1.5pt](2,3)(4,0)
    \uput[-135](0,0){$t$}
    \uput[-45](8,0){$r$}
    \uput[90](4,6){$s$}
    \uput[135](2,3){$a$}
    \uput[-90](4,0){$b$}
    \uput[90](2,1){$bat$}
    \uput[90](4.5,0.5){$bar$}
    \uput[90](4.5,3){$ars$}
    \end{pspicture}
  \end{center}
  \caption{Computing the nets $\s{N}bat$, $\s{N}bar$ and $\s{N}ars$ from $\s{N}rst$}
%\end{example}
\end{figure}
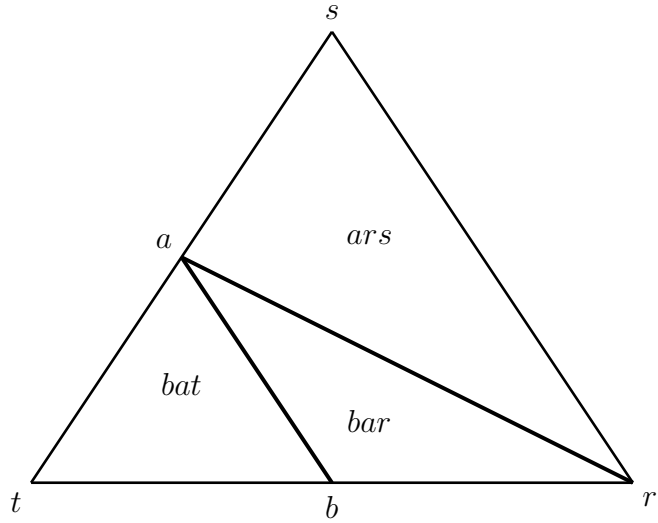

\medskip
We will now compute the net $\s{N}cas$ from the net 
$\s{N}ars$. For this, we need the barycentric coordinates of
$c$ with respect to the triangle $\Delta ars$, which turns out
to be $(0, 1/2, 1/2)$. Using the function {\tt subdecas3sa\/}, the net
$\s{N}cas$ is obtained.

\medskip

\begin{figure}[H]
%\begin{example}
  \begin{center}
    \begin{pspicture}(0,-0.5)(8,6)
    \psline[linewidth=1pt](0,0)(4,6)
    \psline[linewidth=1pt](0,0)(8,0)
    \psline[linewidth=1pt](8,0)(4,6)
    \psline[linewidth=1.5pt](2,3)(6,3)
    \psline[linewidth=1.5pt](2,3)(4,0)
    \psline[linewidth=1.5pt](2,3)(8,0)
    \uput[-135](0,0){$t$}
    \uput[-45](8,0){$r$}
    \uput[90](4,6){$s$}
    \uput[135](2,3){$a$}
    \uput[-90](4,0){$b$}
    \uput[45](6,3){$c$}
    \uput[90](2,1){$bat$}
    \uput[90](4,4){$cas$}
    \uput[90](4.5,0.5){$bar$}
    \end{pspicture}
  \end{center}
  \caption{Computing the net $\s{N}cas$ from $\s{N}ars$}
%\end{example}
\end{figure}
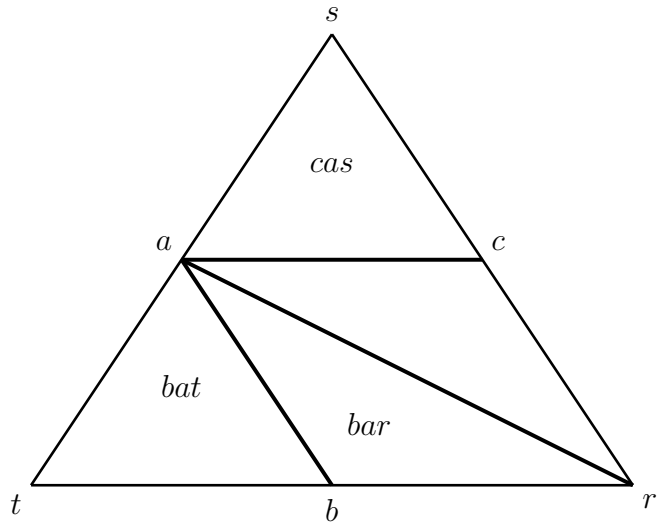

\medskip
We can now compute the nets $\s{N}cbr$ and $\s{N}cba$ from
the net $\s{N}bar$. For this, we need the barycentric coordinates
of $c$ with respect to the affine frame  $\Delta bar$
which turns out to be $(-1, 1, 1)$.
Using the function {\tt sdecas3\/}, the snet
$\s{N}cbr$, $\s{N}car$, and $\s{N}cba$ are obtained, and
we throw away $\s{N}car$.

\medskip

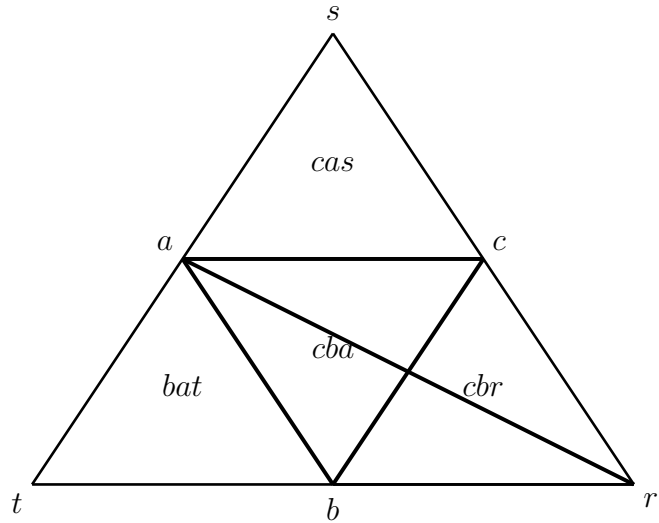
\begin{figure}[H]
%\begin{example}
  \begin{center}
    \begin{pspicture}(0,-0.5)(8,6)
    \psline[linewidth=1pt](0,0)(4,6)
    \psline[linewidth=1pt](0,0)(8,0)
    \psline[linewidth=1pt](8,0)(4,6)
    \psline[linewidth=1.5pt](2,3)(6,3)
    \psline[linewidth=1.5pt](6,3)(4,0)
    \psline[linewidth=1.5pt](2,3)(4,0)
    \psline[linewidth=1.5pt](2,3)(8,0)
    \psline[linewidth=1pt](4,0)(8,0)
    \uput[-135](0,0){$t$}
    \uput[-45](8,0){$r$}
    \uput[90](4,6){$s$}
    \uput[135](2,3){$a$}
    \uput[-90](4,0){$b$}
    \uput[45](6,3){$c$}
    \uput[90](2,1){$bat$}
    \uput[90](4,1.5){$cba$}
    \uput[90](6,1){$cbr$}
    \uput[90](4,4){$cas$}
    \end{pspicture}
  \end{center}
  \caption{Computing the nets $\s{N}cbr$ and $\s{N}cba$ from $\s{N}bar$}
%\end{example}
\end{figure}

\bigskip
Finally, we apply ${\tt transposej\/}$ to the net $\s{N}bat$
to get the net $\s{N}abt$,
${\tt transposek\/}$ to $\s{N}cba$ to get the net $\s{N}bac$,
${\tt transposej\/}$ followed by ${\tt transposek\/}$ to the net $\s{N}cbr$ 
to get the net $\s{N}crb$, and ${\tt transposek\/}$ twice to $\s{N}cas$ 
to get the net $\s{N}sca$,

\begin{figure}[H]
%\begin{example}
  \begin{center}
    \begin{pspicture}(0,-0.5)(8,6)
    \psline[linewidth=1.5pt](0,0)(4,6)
    \psline[linewidth=1.5pt](0,0)(8,0)
    \psline[linewidth=1.5pt](8,0)(4,6)
    \psline[linewidth=1.5pt](2,3)(6,3)
    \psline[linewidth=1.5pt](2,3)(4,0)
    \psline[linewidth=1.5pt](6,3)(4,0)
    \uput[-135](0,0){$t$}
    \uput[-45](8,0){$r$}
    \uput[90](4,6){$s$}
    \uput[135](2,3){$a$}
    \uput[-90](4,0){$b$}
    \uput[45](6,3){$c$}
    \uput[90](2,1){$abt$}
    \uput[90](4,1.5){$bac$}
    \uput[90](6,1){$crb$}
    \uput[90](4,4){$sca$}
    \end{pspicture}
  \end{center}
  \caption{Subdividing $\Delta rst$ into $\Delta abt$, $\Delta bac$,
$\Delta crb$, and $\Delta sca$}
%\end{example}
\end{figure}
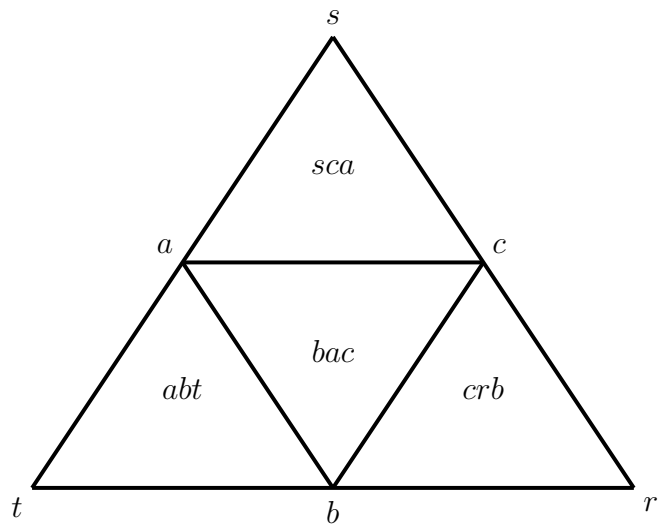

\medskip
Thus, using four calls to the de Casteljau algorithm, we obtained the
nets $\s{N}abt$, $\s{N}bac$, $\s{N}crb$, and $\s{N}sca$.

\remarks 
\begin{enumerate}
\item[(1)] 
For debugging purposes, we assigned different colors
to the patches corresponding to 
$\s{N}abt$, $\s{N}bac$, $\s{N}crb$, and $\s{N}sca$, and
we found that they formed a particularly nice pattern
under this ordering of the vertices of the triangles.
In fact, $\s{N}abt$ is blue, $\s{N}bac$ is red,  
$\s{N}crb$ is green, and $\s{N}sca$ is yellow. 
\item[(2)] 
In the last step of our algorithm, the subdivision step 
is performed with respect to
a point of barycentric coordinates  $(-1, 1, 1)$.
One might worry that such a step involving a nonconvex
combination is a source of numerical instability.
We tested  our algorithm on many different examples, and so far, without
running into any problem. We also believe that such a nonconvex step
is unavoidable if the standard de Casteljau algorithm (building
a simplex of polar values of dimension $3$) is used,
but we are unable to prove this.
\end{enumerate}

\medskip
The subdivision algorithm just presented has been implemented in 
{\it Mathematica\/}, see Gallier \cite{Gallbook}.
The subdivision method is illustrated by the following example
of a cubic patch specified by the control net

\medskip
\begin{verbatim}
 net = {{0, 0, 0}, {2, 0, 2}, {4, 0, 2}, {6, 0, 0}, 
       {1, 2, 2}, {3, 2, 5}, {5, 2, 2}, 
       {2, 4, 2}, {4, 4, 2},  {3, 6, 0}};
\end{verbatim}

\medskip
\begin{figure}[H]
\centerline{
\psfig{figure=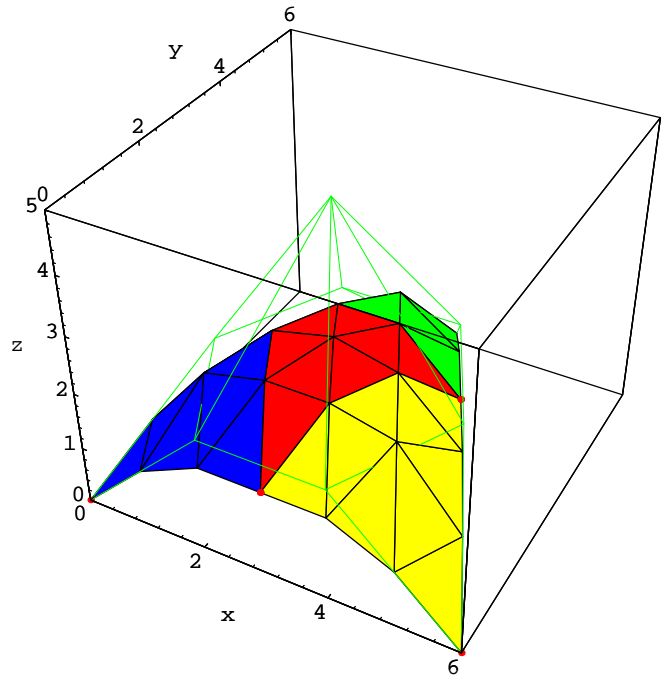,width=3.5truein}}
\caption{Subdivision, $1$ iteration}
\end{figure}

\medskip
\begin{figure}[H]
\centerline{
\psfig{figure=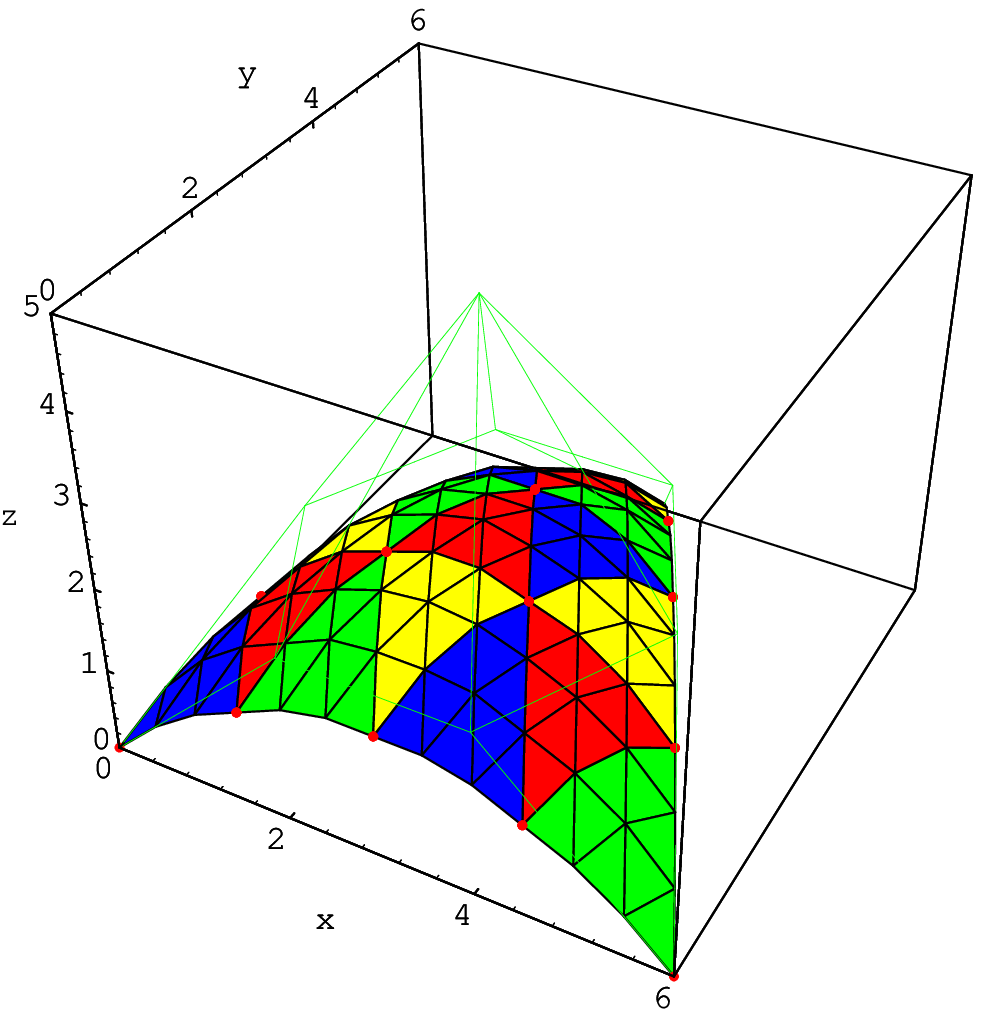,width=3.5truein}}
\caption{Subdivision, $2$ iterations}
\end{figure}

\medskip
\begin{figure}[H]
\centerline{
\psfig{figure=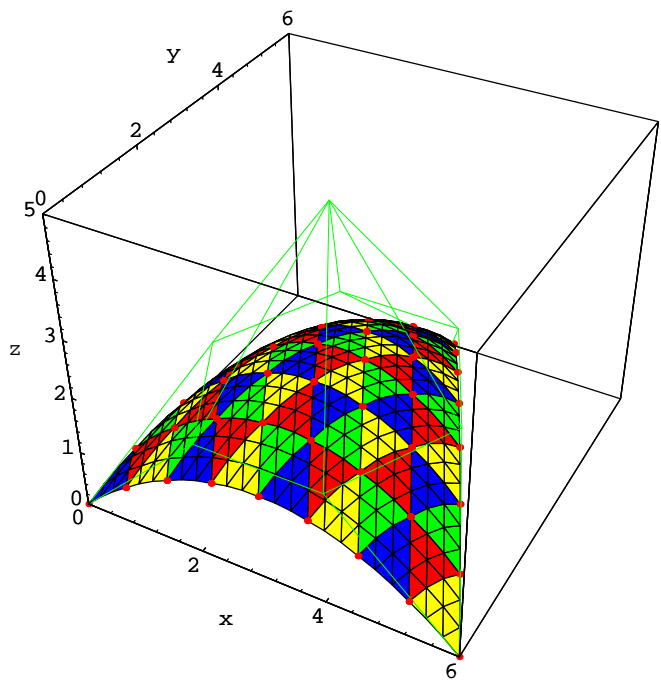,width=3.5truein}}
\caption{Subdivision, $3$ iterations}
\end{figure}

\bigskip
After only three subdivision steps, the triangulation
approximates the surface patch very well.

\medskip
As another example of the use of the above functions,
we can display a portion of a well known
surface known as the ``monkey saddle'',
defined by the equations
$$
x = u,\quad
y = v,\quad
z = u^3 -3uv^2.
$$

%\medskip
Note that $z$ is the real part of the complex number  $(u + iv)^3$. 
It is easily shown that
the monkey saddle is specified by the following 
triangular control net
{\tt monknet\/} over the standard affine frame $\Delta rst$,
where $r = (1,0,0)$, $s = (0, 1, 0)$, and $t = (0, 0, 1)$.

%\medskip
\begin{verbatim}
monknet = {{0, 0, 0}, {0, 1/3, 0}, {0, 2/3, 0}, {0, 1, 0},
          {1/3, 0, 0}, {1/3, 1/3, 0},  {1/3, 2/3, -1},  
          {2/3, 0, 0}, {2/3, 1/3, 0},  {1, 0, 1}};
\end{verbatim}

\medskip
We actually display the patch over the rectangle
$[-1, 1]\times [-1, 1]$. This can be done by splitting the square into two
triangles, and computing control nets with respect to these
triangles. This is easy to do, and it is explained for example in Gallier 
\cite{Gallbook}.
Subdividing both nets $3$ times, we get the following picture.

%\medskip
\begin{figure}[H]
\centerline{
\epsfig{figure=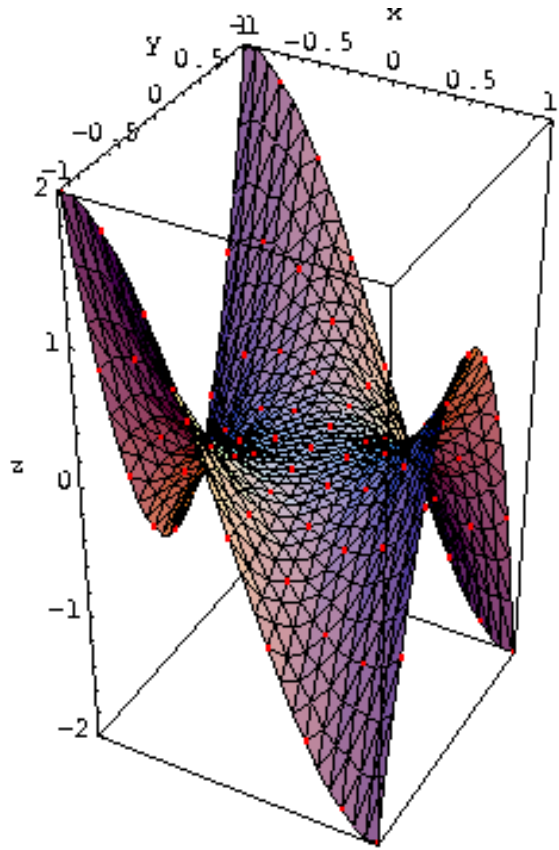,width=4.5truein}}
\caption{A monkey saddle, triangular subdivision}
\end{figure}

\section{A Diamond-Shape Strategy For Subdivision}
\label{sec5}
The strategy of the previous section was to split the affine 
frame $\Delta rts$
into four congruent subtriangles. We were able to do this using
four calls to the de Casteljau algorithm and we showed that
it is not possible to do it in fewer calls.

\medskip
However, it is possible to split the affine frame into 
four subtriangles using only three calls to the de Casteljau algorithm.
The method consists in splitting the triangle
$\Delta rst$ into the four subtriangles 
$\Delta bat$, $\Delta bar$, $\Delta cas$, and $\Delta car$:  

\medskip

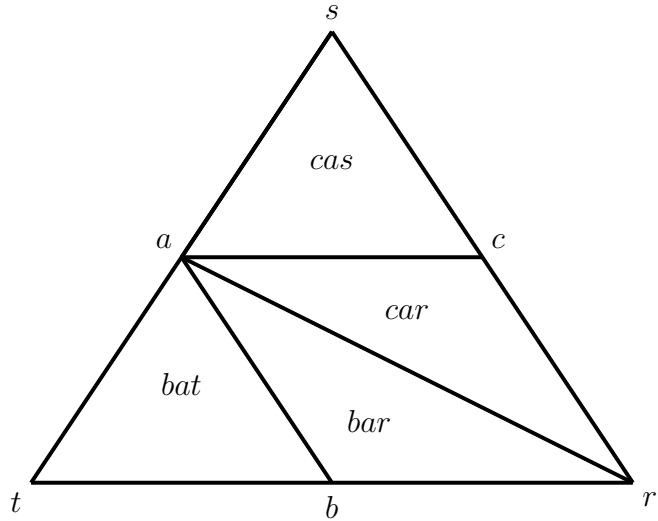
\begin{figure}[H]
%\begin{example}
  \begin{center}
    \begin{pspicture}(0,-0.5)(8,6)
    \psline[linewidth=1.5pt](0,0)(4,6)
    \psline[linewidth=1.5pt](0,0)(8,0)
    \psline[linewidth=1.5pt](2,3)(6,3)
    \psline[linewidth=1.5pt](2,3)(4,0)
    \psline[linewidth=1.5pt](2,3)(8,0)
    \psline[linewidth=1.5pt](2,3)(4,6)
    \psline[linewidth=1.5pt](8,0)(4,6)
    \uput[-135](0,0){$t$}
    \uput[-45](8,0){$r$}
    \uput[90](4,6){$s$}
    \uput[135](2,3){$a$}
    \uput[-90](4,0){$b$}
    \uput[45](6,3){$c$}
    \uput[90](2,1){$bat$}
    \uput[90](4,4){$cas$}
    \uput[90](4.5,0.5){$bar$}
    \uput[90](5,2){$car$}
    \end{pspicture}
  \end{center}
  \caption{Diamond-style subdivision of an affine frame $\Delta rst$}
%\end{example}
\end{figure}

\medskip
This can be done by first computing the nets $\s{N}art$ and $\s{N}ars$,
which can be done in one call to {\tt sdecas3\/} (dropping  $\s{N}ast$).
Next,  we split $\Delta art$ into
the two triangles $\Delta bat$ and $\Delta bar$. For this, we need
the barycentric coordinates of $b$ with respect to the triangle 
$\Delta art$, which turns out $(0, 1/2, 1/2)$.
Using the function {\tt sdecas3\/}, the nets
$\s{N}bat$, $\s{N}brt$, and $\s{N}bar$ are obtained, and we throw away
$\s{N}brt$.
Finally, we split $\Delta ars$ into
the two triangles $\Delta cas$ and $\Delta car$. For this, we need
the barycentric coordinates of $c$ with respect to the triangle 
$\Delta ars$, which turns out $(0, 1/2, 1/2)$.
Using the function {\tt sdecas3\/}, the nets
$\s{N}cas$, $\s{N}crs$, and $\s{N}car$ are obtained, and we throw away
$\s{N}crs$.

\medskip
An implementation of the method is given in Gallier \cite{Gallbook}.

\bigskip
The result of subdividing two of three times reveals some
diamond-shape subdividion patterns.
For example, after three iterations, the dome surface is subdivided as follows:

%\medskip
\begin{figure}[H]
\centerline{
\epsfig{figure=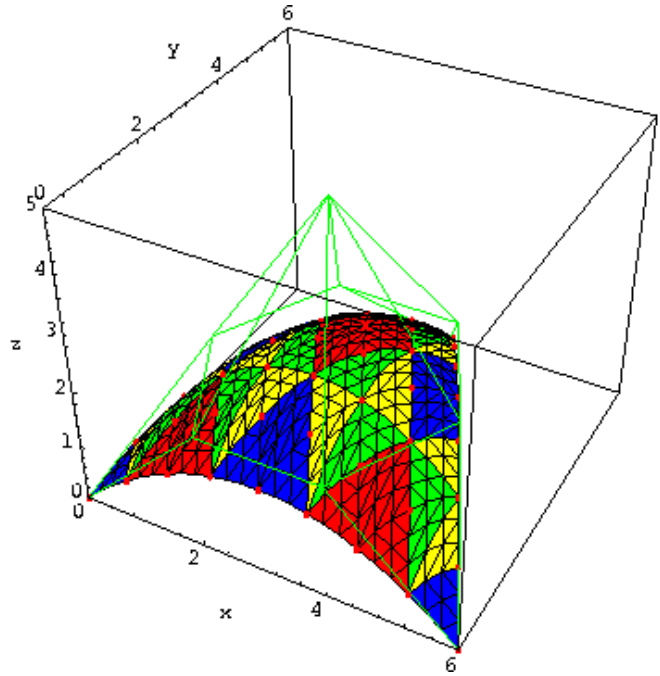,width=3.5truein}}
\caption{Diamond-style subdivision, $3$ iterations}
\end{figure}

\section{A Spider-Web Strategy For Subdivision}
\label{sec6}
Is is also possible to split the affine frame into 
six subtriangles using only four calls to the de Casteljau algorithm.

%\medskip

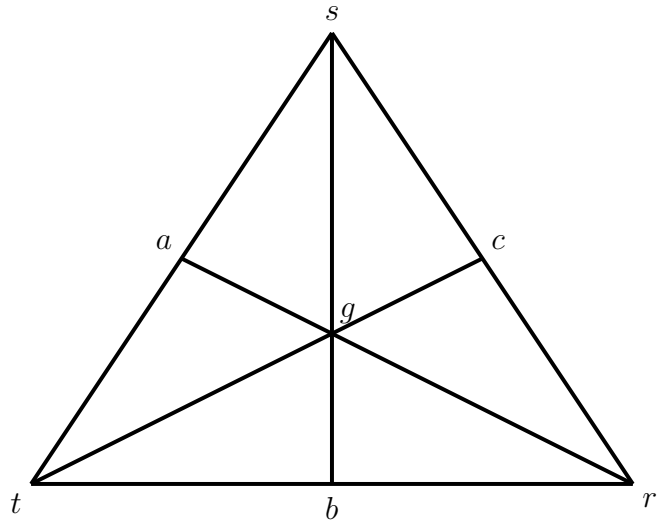
\begin{figure}[H]
%\begin{example}
  \begin{center}
    \begin{pspicture}(0,-0.3)(8,5.5)
    \psline[linewidth=1.5pt](0,0)(4,6)
    \psline[linewidth=1.5pt](0,0)(8,0)
    \psline[linewidth=1.5pt](8,0)(4,6)
    \psline[linewidth=1.5pt](0,0)(4,2)
    \psline[linewidth=1.5pt](8,0)(4,2)
    \psline[linewidth=1.5pt](4,2)(4,6)
    \psline[linewidth=1.5pt](4,2)(2,3)
    \psline[linewidth=1.5pt](4,2)(4,0)
    \psline[linewidth=1.5pt](4,2)(6,3)
    \uput[-135](0,0){$t$}
    \uput[-45](8,0){$r$}
    \uput[90](4,6){$s$}
    \uput[135](2,3){$a$}
    \uput[-90](4,0){$b$}
    \uput[45](6,3){$c$}
    \uput[50](4,2){$g$}
    \end{pspicture}
  \end{center}
  \caption{spider-web subdivision of an affine frame $\Delta rst$}
%\end{example}
\end{figure}
%

%\medskip
The triangle $\Delta rst$ is subdivided into the triangles
$\Delta bgt$, $\Delta bgr$, $\Delta agt$, $\Delta ags$, $\Delta cgs$, and $\Delta cgr$,
as shown above, where $g$ is the center of gravity.
This can be done by first computing the nets $\s{N}grt$, $\s{N}gst$, and  $\s{N}grs$,
which can be done in one call to {\tt sdecas3\/}.
We split $\Delta grt$ into
the two triangles $\Delta bgt$ and $\Delta bgr$ using {\tt sdecas3\/}
(throwing away $\Delta brt$). 
We split $\Delta gst$ into
the two triangles $\Delta agt$ and $\Delta ags$ using {\tt sdecas3\/}
(throwing away $\Delta ast$). Finally, we split $\Delta grs$ into
the two triangles $\Delta cgs$ and $\Delta cgs$ using {\tt sdecas3\/}
(throwing away $\Delta crs$). 

\medskip
The result of subdividing recursively yields spider-web like patterns.
For example, after three iterations, the dome surface is subdivided as follows:

%\medskip
\begin{figure}[H]
\centerline{
\epsfig{figure=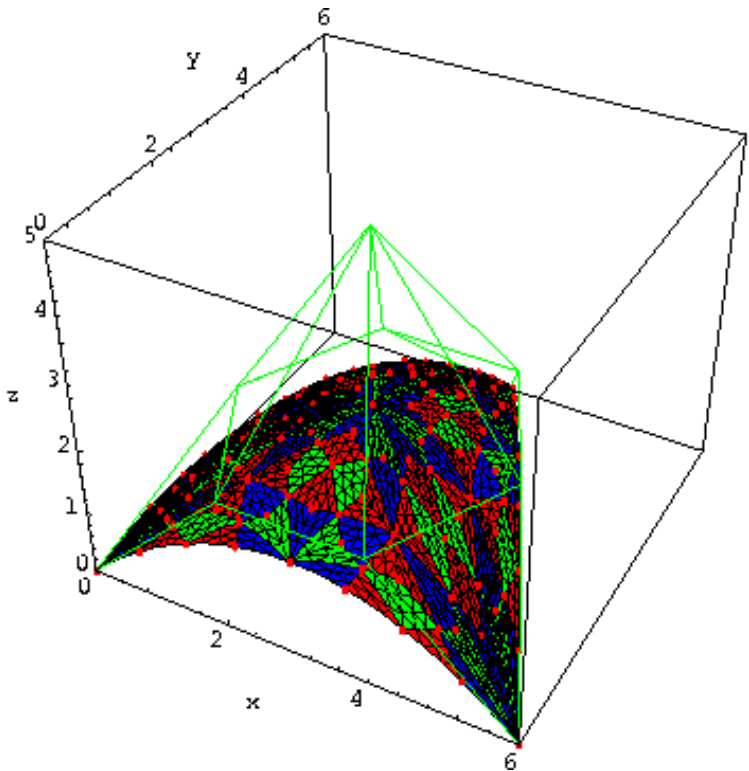,width=4truein}}
\caption{Spider-web style subdivision, $3$ iterations}
\end{figure}

\section{Conclusion}
\label{sec7}
We have presented various strategies for subdividing polynomial triangular surface
patches.  We gave an algorithm performing a regular subdivision
in four calls to the standard de Casteljau algorithm, and
we showed  that this method for obtaining a regular subdivision
is optimal. We gave another  subdivision algorithm using only three calls 
to the de Casteljau algorithm. Instead of being regular,
the subdivision pattern is  diamond-like. Finally, we presented a
``spider-web'' pattern subdivision scheme producing six subtriangles in
four calls to the de Casteljau algorithm.
These methods immediately
apply to rational surface patches (Gallier \cite{Gallier96c}).

\medskip
An amusing effect is obtained from the regular subdivision scheme  if we omit
the central triangle $\Delta bac$. We obtain a ``fractalized'' representation of
the surface patch, in the sense that a Sierpinski gasket pattern is
laid onto the patch! It would be interesting to investigate other
subdivision strategies and the patterns that they induce, especially if
the triangles are colored in various recursive manners.

\bibliography{../basicmath/cadgeom}
\bibliographystyle{plain} 

\end{document}